\documentclass[final,3p,times]{elsarticle}

\usepackage{epsfig}
\usepackage{amssymb, amsmath}
\usepackage{amsthm}
\usepackage{graphicx}
\usepackage{dcolumn}
\usepackage{bm}
\usepackage{widetext}
\usepackage{flushend}
\usepackage{cuted}

\begin{document}

\begin{frontmatter}



\title{A simple metal-insulator criterion for the doped Mott-Hubbard materials}


\author{Vladimir A. Gavrichkov}

\address{L. V. Kirensky Institute of Physics, Siberian Branch of Russian Academy of Sciences, 660036, Krasnoyarsk, Russia}

\begin{abstract}
A simple metal-insulator criterion for doped Mott-Hubbard materials has been derived. Its readings are closely related to the orbital and spin nature of the ground states of the unit cell. The available criterion readings (metal or insulator) in the paramagnetic phase reveal the possibility of the insulator state of doped materials with the forbidden first removal electron states. According to its physical meaning, the result is similar to the Wilson's criterion in itinerant electron systems. The application of the criterion to high-$T_c$ cuprates is discussed.

\end{abstract}


\end{frontmatter}


\section{Introduction}
It is known that some of perspective doped transition-metal oxides: cuprates
~\cite{Kohsaka_etal2008, Yang_etal2008} and manganites ~\cite{Saitoh_etal2000} show mysterious pseudogap states.
It is surprising that the pseudogap states in such different oxides have common features~\cite{Rui_etal2014, Mannella_etal2005}. Extensive dielectric regions on the phase $T/x$-diagram of manganites~\cite{Cheong_etal1999, Coey_etal1999} also attract attention. (primarily in the paramagnetic phase(PM)). If the reasons for the insulator states in these different doped oxides are general by nature, one can try to detect them  directly following the quasiparticle picture established by J. Hubbard~\cite{Hubbard_1963}.

The purpose of our work is to construct a metal-insulator criterion based on Wilson's ideas~\cite{Wilson_1931} concerning a system of itinerant electrons in the analytical form for the
doped Mott-Hubbard materials.
Indeed, this approach includes the statement that if the electron system consists of completely occupied
and empty bands, it is an insulator, otherwise, it is a metal.
Here, there are some features. Due to many-electron effects the spectral density of quasiparticle states in the Mott-Hubbard materials
depends on the carrier concentration. Secondly, it does not make much sense to apply the Wilson's criterion to the doped materials because of the fluctuations of the impurity potential which create new states.
However, it is known that because of the similarity of the phase $T/x$ diagrams of the high-T$_c$ cuprates,
with the carriers having different origins, the carrier concentration is a crucial factor. Consideration is given to the criterion taking into account the many-electron effects only.

\section{Method}
 The generalized approach uses the fact that the optical intracell transitions with their ($l$-orbital, $S$-spin)-selection rules in the transparency window and optical charge transfer transitions in the oxides can be observed at the same $d$-states~\cite{Krinchik_etal1969, Eremenko_etal1969}.
In the first approximation one can assume that the quasiparticles are unit cell excitations which can be represented graphically as single-particle transitions between different sectors $N_h=...(N_{h0}-1), N_{h0}, (N_{h0}+1),...$ of the configuration space of the unit cell ($N_{h0}$-hole number per cell in the undoped material, see Fig.\ref{fig:1})~\cite{Ovchinnikov_etal2012}. Each of these transition forms a $r$-th quasiparticle band, where the root vector $r=\{ii'\}$ in the configuration space numerates the initial $i$ and final $i'$ many-electron states in the transition. The transitions, with the number of electrons increasing or decreasing, form the conduction or valence bands, respectively.

For our purposes it is convenient to start with Lehmann's representation for the Green's function $G_{fg\sigma} ^{\lambda\lambda}$
of the intracell Hamiltonian $H_0$  with respect to the family of single-particle operators  $c_{f\lambda\sigma}^{(+)}$ and their matrix elements in the basis of ${\left| {({N_{h}},{M_S})_i} \right\rangle} $ - eigenstates of $H_0$  ($S$ and $M$ are the spin and spin projection of the multi-electron cell eigenstate):
\begin{eqnarray}
G_{fg\sigma}^{\lambda\lambda}&=&\langle\langle c_{f\lambda\sigma}|c^+_{g\lambda\sigma}\rangle\rangle=
\sum\limits_{rr'} ^{}\gamma_{f\lambda\sigma}(r)\gamma_{g\lambda\sigma}(r')D_{0fg}^{rr'}(E)=\nonumber\\
 &=& {\delta _{fg}}\sum\limits_{rr'} ^{}{\delta _{rr'}}\frac{{\gamma_{\lambda\sigma}^2(r)F_r\left( x \right)}}{{E - {\Omega _r}}},
\label{eq:1}
\end{eqnarray}
where matrix elements
\begin{eqnarray}
 {\gamma _{\lambda \sigma }}\left( {{r}} \right)& = &\left\langle ({{N_{h} + 1,{{M'}_{S'}}})_\tau} \right|c_{f\lambda \sigma } \left|({{N_{h},{M_S}})_\mu}\right\rangle\times \nonumber  \\
 & \times &\delta \left( {S',S \pm |\sigma| } \right)\delta \left( {M', M + \sigma } \right),
\label{eq:2}
\end{eqnarray}
the total space of the root vectors $\{r\}=...+\{r_{12}\}+\{r_{23}\}+..., (\{r_{12}\}=\{\mu\tau\},\{r_{23}\}=\{\tau\eta\}$ and so on, Fig.\ref{fig:1}). An occupation factor $F_r\left( x \right)$ is the probability to detect a cell in any of the $i,i'$ states participating in the $r$-th transition, and
$\Omega _r=E_i({N_{h}},{M_S})-E_{i'}({N_{h}+1},{{M'}_{S'}})$ is a quasiparticle energy in the $r$-th band. For example, in the PM phase of the doped material the occupation factor has the  form:
\begin{equation}
F_{r_{12}}\left( x \right) = \frac{{1 - \alpha x}}{{2S + 1}},
\end{equation}
where $\alpha=1-({2S+1})/({2S'+1})$ is proportional to the ratio of the spin multiplets of the $i,i'$ states participating in the $r_{12}$ - (from the subspace $\{r_{12}\}$) transition between the ground states ${\left| {({N_{h0}},{M_S})_{i=0}} \right\rangle} $ and ${\left| {({N_{h0}+1},{M'_{S'}})_{i'=0}} \right\rangle}$ indicated by the arrow in Fig.\ref{fig:1}

The Green's function in its simplest form (\ref{eq:1}) is quite insufficient to study the $\vec{k}$-physics of quasiparticles (eg. superconductivity properties  of high-${T_c}$ cuprates).  However, this approach is free from the shortcomings of the hydrogen-like (s-)representation and low-energy approximations, because we do not restrict ourselves
to choosing the intracell Hamiltonian $H_0$, and we are ready to work with all   ${\left| {({N_{h}},{M_S})_i} \right\rangle}$
states in the framework of the Russell-Saunders scheme. The total number of the valence states is equal
to the sum over all the quasiparicle valence states:

\begin{widetext}
\begin{equation}
N_v(x)= {\sum\limits_{\lambda\sigma}  {\sum\limits_{r}
{{\gamma _{\lambda \sigma }}^2\left( r \right)}
{{ \int\limits_{}  {dE\left( { - \tfrac{1}{\pi }}\right){\operatorname{Im} {D_0^{r}}\left( {E} \right)}}_{E + i0}}} }} =N_v^{12}(x)+N_v^{23}(x),
\label{eq:3}
\end{equation}
\end{widetext}
where $N_v^{12}(x)$ and $N_v^{23}(x)$ are the contributions from the quasiparticles with the root vectors $r$ from the $\{r_{12}\}$ and  $\{r_{23}\}$ subspaces because the other states of ${\left| {({N_{h}},{M_S})_i} \right\rangle} $ in the doped material are not occupied,
and there is a zero probability $F_r\left( x \right) =0$ to detect a cell in these states at a low temperature. The key condition at the insulating state, which we are interested in, is
\begin{equation}
N_e-x=N_v(x),
\label{eq:4}
\end{equation}
where $(N_e-x)$ is the  total electron number per cell of the hole doped material. That is, if the number of electrons in a cell equals to the number of the valence states, the material is an insulator.
\begin{figure}
\includegraphics{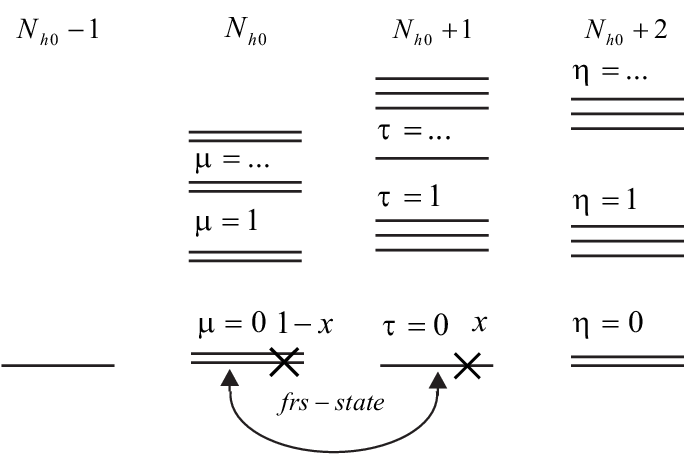}%
\caption{\label{fig:1} $E_i({N_{h}},{M_S})$ - energy level scheme of the Hilbert space based on cell states (\ref{eq:6}) and (\ref{eq:7}) with hole numbers per cell $N_h=N_{h0}-1, N_{h0}, N_{h0}+1,..$, where $i=\mu, \tau, \eta$ and $N_{h0}$ is a hole number per cell in the undoped material. The cross indicates the occupied ground cell states of hole doped material. A solid line with arrows corresponds to the first removal electron states with a lowest binding energy.}
\label{fig:1}
\end{figure}

To obtain the Fermi level position in the degenerate doped material at zero temperature
one could carry out the integration on the right side of the equation
\begin{equation}
x={\sum\limits_{\lambda\sigma}  {\sum\limits_{r}
{{\gamma _{\lambda \sigma }}^2\left( r \right)}
{{ \int\limits_{E_F}  {dE\left( { - \tfrac{1}{\pi }}\right){\operatorname{Im} {D_0^{r}}\left( {E} \right)}}_{E + i0}}} }},
\label{eq:99}
\end{equation}
over the top valence band of the first removal electron states ($frs$) with the lowest binding energy (see Fig.\ref{fig:1}), and this is sufficient at the actual concentrations $x \sim 0.1$, as a rule.  However, this is not sufficient, when the hole concentration $x$ exceeds the number of quasiparticle states in the top valence band $x \gg {N_{frs}}$. And this is not due to the hole concentration being too large, but because the number of $frs$ quasiparticle states $N_{frs}$ may be very small, which is to be seen later. It is not obvious that the Fermi level with doping will be immersed into the following energy valence band in the doped Mott-Hubbard material. To understand the features of the solutions of equation (\ref{eq:99}) in the doped materials with the forbidden $frs$ quasiparticle states (i.e. ${N_{frs}} = 0$),
one derives the total number of the valence states in (\ref{eq:4}) as a function ${N_v}\left( {x,{N_{frs}}} \right)$.

By following this approach, one obtains a simple metal-insulator criterion, which is characterized by the condition:
$N_{frs}=0$ (-insulator) or $N_{frs}\neq 0$ (-metal) irrespective of the hole concentration $x$.
Consideration is given to the case with one hole per cell $N_{h0}=1$ in the undoped materials and an arbitrary number $N_\lambda$ of the occupied $\lambda$ orbitals, i.e. $N_e=2N_\lambda-1$. This is relevant for the high-$T_c$ cuprates.
In this case of one hole per cell, the ${\left| {({N_{h}},{M_S})_i} \right\rangle} $ cell states are a superposition of different hole configurations of the same orbital (l-)symmetry:
\begin{equation}
\left| ({N_{h0},{M_{S}}})_{\mu} \right\rangle  = \sum\limits_\lambda ^{} {{\beta _\mu }\left( {{h_\lambda }} \right)\left| {{h_\lambda },{M_{S}}} \right\rangle }
\label{eq:5}
\end{equation}
Thus, there are one-hole spin doublet states, $C_{{2N_\lambda}}^{1} = {2N_\lambda}$, where  $C^k_n$ is the number of combinations. Besides, there are  $C^2_{2N_\lambda}=N_S+3N_T$ of the spin singlets $N_S=C_{N_\lambda}^{2}+N_\lambda$ and triplets $N_T=C_{N_\lambda}^{2}$:
\begin{equation}
\left| ({N_{h0}+1,{M'_{S'}}})_{\tau} \right\rangle  = \sum\limits_{\nu\nu'} {{B_\tau }\left( {{h_\nu},{h_{\nu'}}} \right)\left| {{h_\nu},{h_{\nu'}},{M'_{S'}}} \right\rangle }
\label{eq:6}
\end{equation}
in the two-hole sector (Fig.\ref{fig:1}) in the ${N_\lambda}$ -orbital approach. Using the intracell Hamiltonian $H_0$ in the cell function  representation
the configuration weights $\beta_\mu(h_\lambda)$ and $B_\tau(h_\lambda,h_{\lambda'})$ can be obtained by the exact diagonalization procedure  for the matrices $(\hat{H_0})_{\lambda\lambda'}$ and $(\hat{H_0})^{\nu\nu'}_{\lambda\lambda'}$ in the $E_i({N_{h},{M_S}})$-eigenvalue problem  in different sectors $N_h$.~\cite{Ovchinnikov_etal2012} The sum (\ref{eq:3}) over all the $r$-th excited states with $\mu\neq0$ in the  sector $N_h=N_{h0}$ is omitted, and only the excited states with any $\tau$($\eta$) index in the  nearest $N_h=(N_{h0}+1)$ and $(N_{h0}+2)$ sectors are summed up.
To calculate the matrix elements (\ref{eq:2}) the eigenfunctions (\ref{eq:5}) and (\ref{eq:6}) in different sectors
must be expressed through each other. It is possible due to the rules for the addition of the angular momenta ~\cite{Landau_etal1974}. The expressions for high- and low-spin two-hole partners (with $S'=S\pm|\sigma|$) can be combined into a single expression:
\begin{widetext}
\begin{equation}
\left| {h_{\lambda},{h_{\lambda '} ,M'_{S'}}} \right\rangle = \Gamma _ \uparrow ^{}\left( {S'_{M'},S} \right)c_{\lambda ' \downarrow } \left| {{h_\lambda },{M'-\textstyle{1 \over 2}}} \right\rangle  + {\mathop{\rm sgn}} (\Delta S)\Gamma _ \downarrow ^{}\left( {S'_{M'},S} \right)c_{\lambda ' \uparrow} \left| {{h_\lambda },{M'+\textstyle{1 \over 2}}} \right\rangle
\label{eq:7}
\end{equation}
\end{widetext}
where $\Delta S=S'-S=\pm|\sigma|$, and the coefficients
\begin{equation}
\Gamma _\sigma ^2\left( {S'_{M'},S} \right) = \frac{{S + \eta \left( \sigma  \right){\mathop{\rm sgn}} \left( {\Delta S} \right)M' + {\textstyle{1 \over 2}}}}{{2S + 1}}
\label{eq:8}
\end{equation}
have a completeness property for the contributions from the identical spin states of a doped hole to different high- and low-spin two-hole partners:
\begin{equation}
\sum\limits_{\Delta S = - \left| \sigma  \right|}^{ + \left| \sigma  \right|} {\Gamma _\sigma ^2\left( {S'_{M'},S } \right)} = \sum\limits_{\sigma}^{ }{\Gamma _\sigma ^2\left( {S'_{M'},S } \right) = 1},
\label{eq:9}
\end{equation}
and also
\begin{equation}
\sum\limits_{M =  - S}^S {{\Gamma _\sigma ^2\left( {{{S'}_{M'}},S} \right)} }
=S + \frac{1}{2}
\label{eq:10}
\end{equation}
Taking into account relations (\ref{eq:5}), (\ref{eq:6}) and (\ref{eq:10}) one can determine the matrix element in (\ref{eq:3}) by the sum:
\begin{widetext}
\begin{equation}
\left\langle ({{N_{h0}} + 1,{{M'}_{S'}}})_\tau \right|c_{\nu \sigma } \left| ({{N_{h0}},{M_S}})_\mu \right\rangle  = \sum\limits_{\lambda ,\lambda ',\lambda ''} {\left\langle {{h_{\lambda '}},{h_{\lambda ''}},M'_{S'}\left| {c_{\nu \sigma }} \right|{h_\lambda },M_{S}} \right\rangle {\beta _\mu }\left( {{h_\lambda }} \right){B_\tau }\left( {{h_{\lambda '}},{h_{\lambda ''}}} \right)\Gamma _\sigma \left( {S'_{M'} ,S} \right)}
\label{eq:11}
\end{equation}
\end{widetext}
After substituting expression (\ref{eq:11}), (\ref{eq:2}) and Green's function (\ref{eq:1}) into relation (\ref{eq:3}) one obtains:
\begin{equation}
N_v (x)= N_v^{12}+N_v^{23}={N^{12}_{s,v}} + 3{N^{12}_{t,v}}+N_v^{23},
\label{eq:12}
\end{equation}
where instead of the "blind" sum over the root vectors ${{r}}$, we use the summation over the physically meaningful indices $\tau $,  $M$ and $\Delta S$
(i.e. the sum over all low- spin ($s$)  and high-spin ($t$) two-hole states).
\begin{widetext}
\begin{equation}
N_{s(t),v}^{12}= \sum\limits_{\nu \sigma } {\sum\limits_{\tau }
{{F^{s(t)}_{r = \left\{ {0,\tau } \right\}}}\left( x \right)\sum\limits_{MM'}^{}
{{{\left\{ {\sum\limits_\lambda  {\Gamma _\sigma \left( {S'_{M'}, S} \right){\beta _{\mu  = 0}}
\left( {{h_\lambda }} \right)} {B_\tau }\left( {{h_\lambda },{h_\nu }} \right)\delta \left( {S',S \pm |\sigma| } \right)
\delta \left( {M', M +  \sigma} \right)} \right\}}^2}} } },
\label{eq:13}
\end{equation}
\end{widetext}
where the $(+)$ and $(-)$ on the right side are used with the indices $t$ and $s$, respectively, and the occupation factor in the PM phase reads:
\begin{equation}
{F^{s(t)}_{\left\{ {0,\tau } \right\}}}(x) = \left\{ {\begin{array}{*{20}{c}}
   {\tfrac{1}{2}\left( {1 - \alpha_{s(t)} x} \right),\tau  = 0}  \\
   {\tfrac{1}{2}\left( {1 - x} \right),\tau  \ne 0}  \\
\end{array} } \right.,
 \label{eq:14}
\end{equation}
with $\alpha_{s(t)}=1-{2}/{(2S'+1)}$ and $S'=0,1$; $S=1/2$. Let's start with the contribution from the spin singlet $frs$ states $N^s_{frs}$:
\begin{widetext}
\begin{equation}
N_{v}(x) =(2N_\lambda-1)-x(1-N^s_{frs})=N_e-x(1-N^s_{frs})
\label{eq:15}
\end{equation}
\end{widetext}
where the low and high spin contributions are
\begin{equation}
N^{12}_{s,v}=(1/2)[(N_\lambda+1)(1-x)+2xN^{s}_{frs}]
\label{eq:16}
\end{equation}
 and
\begin{equation}
N^{12}_{t,v}=(1/2)(N_\lambda-1)(1-x)
\label{eq:17}
\end{equation}
respectively.
The relation $N_v^{23}=x(2N_\lambda-2)$ for the contributions from the quasiparticle with the root vectors from $\{r_{23}\}$ subspace is derived similar to the previous expression for contribution (\ref{eq:13}).
The number of possible singlet $frs$ states is in the range $0 \leq N^s_{frs} \leq 2$, where
\begin{eqnarray}
N_{frs}^s &=& 1 - \sum\limits_\lambda  {\beta _0^2\left( {{h_\lambda }} \right)}\times\nonumber\\ &&\sum\limits_{\lambda ',\lambda ''} {\left[ {1 - {\delta _{\lambda \lambda '}} - {\delta _{\lambda \lambda "}}} \right]B_{\tau  = 0}^2\left( {{h_{\lambda '}}{h_{\lambda ''}}} \right)},
\label{eq:18}
\end{eqnarray}
and $ \tau  = 0   $ corresponds to the $frs$-quasiparticles. In deriving (\ref{eq:15}) we also use relation (\ref{eq:10}) and identity
\\$\sum\limits_\lambda  {\beta _\mu ^2\left( {{h_\lambda }} \right)} \sum\limits_{\lambda ',\lambda ''} {B_\tau ^2\left( {{h_{\lambda '}}{h_{\lambda ''}}} \right)}  = 1$ at any $\mu$  and $\tau$. Since the sum
\begin{equation}
\sum\limits_\tau  {\left[ {{\beta _0}\left( {{h_\lambda }} \right){B_\tau }\left( {{h_\lambda },{h_\nu }} \right)} \right]\left[ {{\beta _0}\left( {{h_{\lambda '}}} \right){B_\tau }\left( {{h_{\lambda '}},{h_\nu }} \right)} \right]}  = 0
\label{eq:19}
\end{equation}
at any $\nu $ and $\lambda  \ne \lambda '$, the contribution from the cross-term  from (\ref{eq:13}) to the total number of the valence states is absent.
In the case of the triplet nature of $frs$ states, one obtains a similar expression to (\ref{eq:15}) with the
contribution
\begin{equation}
N_{frs}^t = 1 - \sum\limits_\lambda  {\beta _0^2\left( {{h_\lambda }} \right)} \sum\limits_{\lambda ' \ne \lambda '' \ne \lambda } {B_{\tau  = 0}^2\left( {{h_{\lambda '}}{h_{\lambda ''}}} \right)},
\label{eq:20}
\end{equation}
where $0 \le {N_{frs}^t} \le 1$.
\section{Results and discussion}
From relation (\ref{eq:15}) it follows that the doped material can show both the metallic $N_v(x)>(N_e-x)$, and dielectric properties $N_v(x)=(N_e-x)$ at $N^{s(t)}_{frs}>0$ or $N^{s(t)}_{frs}=0$, respectively. This result is similar to the Wilson's criterion for the itinerant electron systems~\cite{Wilson_1931}.
The $frs$ states can be prohibited at $\delta (S',S\pm|\sigma|)=0$ (the s-forbidden $frs$ states)  as well as when the doped hole changes the initial orbital configuration of $\left| ({N_{h0},{M_{S}}})_{\mu=0} \right\rangle$ ground cell states as a whole (l-forbidden $frs$ states). Surprisingly, only one forbidden transition in the cell leads to the insulating state of the whole material. How this transition is different from many others? Nothing changes in the undoped material. The transition is distinguished only in the doped material by a specific symmetry of the ground states of doped carriers.

Under the conditions $N_\lambda=1$ and $N_e=(1-x)$ one  always obtains  a simple metal with the $N^s_{frs}=2$ and $N_v(x)=(1+x)$ valence states,
as in the Hubbard model, where the high-spin (triplet) states are simply not available.
\begin{figure}
\includegraphics{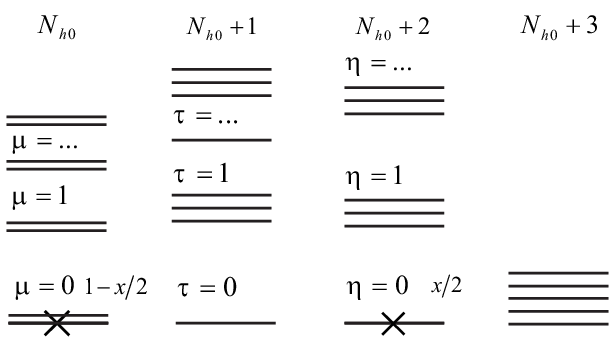}
\caption{\label{fig:2} }
\end{figure}
Note, from the derivation of expression (\ref{eq:15}) it is clear that the orbital and spin degeneracy of the ground cell state in the sector $N_h=N_{h0}$ does not result in the metallic character of the undoped  material.

To apply the criterion to the high-$T_c$ cuprates we choose the intracell Hamiltonian $H_0$ in accordance with the LDA+GTB approach~\cite{Ovchinnikov_etal2012}. This method generalizes the Hubbard's concept to real materials by using the Wannier cell function representation~\cite{Shastry_1989, Feiner_etal1996, Hayn_etal1999, Gavrichkov_etal2001}. Let's calculate the magnitude of $N^{s(t)}_{frs}$ in the high-$T_c$ cuprates, where $r=\{^2b_1,A_1\}$ - root vector which is  relevant for (\ref{eq:13}) at $\mu=0$ and $\tau=0$~\cite{Feiner_etal1996, Gavrichkov_etal2001}, i.e. it corresponds only to the $A_1$ singlet $frs$ state.
Using the exact diagonalization procedure with LDA parameters~\cite{Korshunov_etal2005}, one obtains the relation:
\begin{eqnarray}
N_{frs}^s&=&1+[\beta^2_0(h_b)-\beta^2_0(h_{d_x})]\times \nonumber \\
&\times&[B^2_0(h^2_b)-B^2_0(h^2_{d_x})]\approx0.97
\label{eq:21}
\end{eqnarray}
for the singlet $frs$ states, where doublet and singlet ground states (\ref{eq:5}) and (\ref{eq:6}) are
\begin{eqnarray}
|^2b_1\rangle_0&=&\beta_0(h_b)|h_b,\sigma_{\frac{1}{2}}\rangle+\beta_0(h_{d_x})|h_{d_x},\sigma_{\frac{1}{2}}\rangle \nonumber \\
|A_1\rangle_0&=&B_0(h^2_b)|h^2_b,0_0\rangle+B_0(h^2_{d_x})|h^2_{d_x},0_0\rangle+ \nonumber \\
&+&B_0(h_{d_x},h_b)|h_{d_x},h_b,0_0\rangle,
\label{eq:22}
\end{eqnarray}
$h_b$ and $h_{d_x}$ are the holes in the $b$-symmetrized cell states of oxygen and $d_{x^2-y^2}$ cooper states of the CuO$_2$ layer, respectively. There are no forbidden states, and the number of the valence states is almost constant: $N_v(x)\approx N_e-0.03x$.
However, if the doped holes for any physical reasons, avoid the states of the sector $N_h=(N_{h0}+1)$ and all of them are in the ground cell state as the hole pairs in the sector $(N_{h0}+2)$ (see Fig.\ref{fig:2}), one has  $N_v(x)=(1-x/2)(2N_\lambda-1)+(2N_\lambda-3)(x/2)=N_e-x$,
and, therefore, the material should be an insulator.
In fact, at $T>T_c$ in the high-$T_c$ cuprates there is a mysterious pseudogap state
not associated with the fluctuations of the superconducting order~\cite{Kohsaka_etal2008, Yang_etal2008, Rui_etal2014}.
The number of the $frs$ states $N_{frs}$ is determined by the ability to correctly describe
the interaction of the doped holes in the intracell Hamiltonian $H_0$.
In particular, to numerically evaluate the number of the valence states in (\ref{eq:21})
it is possible to choose an approach different from LDA.

We also believe that it is necessary to investigate doped manganites with the high temperature pseudogap in the PM phase~\cite{Saitoh_etal2000} and cobaltites with the spin forbidden $frs$ states $\delta (S',S\pm|\sigma|)=0$ $(S=0$ and $S'=2)$  at the Co$^{3+}$ and Co$^{4+}$ ground states  ~\cite{Orlov_etal2013} in connection with the subject under discussion.
\section{Conclusions}
Formally, the criterion $N_{frs}=0$ is valid at any hole concentration $x$ and, unlike the Mott-Hubbard transition, a band crossover can occur at the ratio $t_{\lambda\lambda'}/U\ll1$ between the hopping and the Coulomb interaction $\sim U(C_{N_{h0}+1}^{2}+C_{N_{h0}-1}^{2}-2C_{N_{h0}}^{2})$. A source of the metal-insulator transition could be any external effect resulting in the crossover between the top valence band and the forbidden $frs$ state level, with  the insulator becoming a metal. And we don't expect any effects from new quasiparticle states~\cite{Georges_etal1996}  in this range of Hamiltonian's parameters.

\section*{Acknowledgments}
We acknowledge with pleasure discussions with Igor S. Sandalov during the course of this work. This work was supported by RFBR Grant No.13-02-01395, No.14-02-00186, NSh-1044.2012.2 and UrB-SB RAS Grant No.44.
\bibliographystyle{elsarticle-num}
\bibliography{my}

\end{document}